\documentclass[pra,onecolumn]{revtex4-1}
\usepackage{amssymb}
\usepackage{amsmath}
\usepackage{graphicx}
\usepackage{subfigure}
\usepackage{epstopdf}
\usepackage{color}

\begin{document}

\title{Spontaneous Symmetry Breaking in Nonlinear Systems: an Overview and a
Simple Model}
\author{Boris A. Malomed}
\affiliation{Department of Physical Electronics, School of Electrical Engineering,
Faculty of Engineering, Tel Aviv University, Tel Aviv 69978, Israel}

\begin{abstract}
The paper combines two topics belonging to the general theme of the
spontaneous symmetry breaking (SSB) in systems including two basic competing
ingredients: the self-focusing cubic nonlinearity and a
double-well-potential (DWP) structure. Such systems find diverse physical
realizations, chiefly in optical waveguides, made of a nonlinear material
and featuring a transverse DWP structure, and in models of atomic BEC with
attractive inter-atomic interactions, loaded into a pair of symmetric
potential wells coupled by tunneling across the separating barrier. With the
increase of the nonlinearity strength, the SSB occurs at a critical value of
the strength. The first part of the paper offers a brief overview of the
topic. The second part presents a model which is designed as the simplest
one capable to produce the SSB phenomenology in the one-dimensional
geometry. The model is based on the DWP built as an infinitely deep
potential box, which is split into two wells by a delta-functional barrier
at the central point. Approximate analytical predictions for the SSB are
produced for two limit cases: strong (deep) or weak (shallow) splitting of
the potential box by the central barrier. Critical values of the strength of
the nonlinearity at the SSB point, represented by the norm of the stationary
wave field, are found in both cases (the critical strength is small in the
former case, and large in the latter one). For the intermediate case, a less
accurate variational approximation (VA) is developed.
\end{abstract}

\maketitle

\section{Introduction and an Overview of the Topic}

Properties of collective excitations in physical systems are determined by
the interplay of several fundamental ingredients, \textit{viz}., spatial
dimension, external potential acting on the corresponding physical fields\
(or wave functions), the number of independent components of the fields, the
underlying dispersion relation for linear excitations, and, finally, the
character of the nonlinear interactions of the fields. In particular, the
shape of the external potentials determines the system's symmetry, two most
common types of which correspond to periodic (alias \textit{lattice})
potentials and double-well potentials (DWPs), the latter featuring the
symmetry between two wells separated by a potential barrier. The well are
coupled by the tunneling of fields across the barrier, which is an
essentially linear effect.

One of fundamental principles of quantum mechanics (that, by itself, is a
strictly linear theory) is that the ground state (GS) of the quantum system
exactly follows the symmetry of the potential applied to the system. On the
other hand, excited states may realize other representations of the same
symmetry \cite{LL}. In particular, the GS wave function for a quantum
particle trapped in the one-dimensional DWP is symmetric, i.e., even, with
respect to the double-well structure, while the first excited state always
features the opposite parity, being represented by an antisymmetric
(spatially odd) function. A similarly feature of Bloch wave functions
supported by periodic potentials is that the state at the bottom of the
corresponding lowest Bloch band features the same periodicity, while generic
Bloch functions are quasi-periodic ones, with the quasi-periodicity
determined by the quasi-momentum of the excited states.

While the quantum-mechanical Schr\"{o}dinger equation is linear for the
single particle, the description of ultracold rarefied gases formed by
bosonic particles (i.e., the Bose-Einstein condensate, BECs) is provided by
the Gross-Pitaevskii equation (GPE), which, in the framework of the
mean-field approximation, takes into regard effects of collisions between
the particles, by means of an cubic term added, to the Schr\"{o}dinger
equation for the single-particle wave function \cite{BEC}. The repulsive or
attractive forces between the colliding particles are accounted for by,
respectively, the self-defocusing, alias self-repulsive, or self-focusing,
i.e., self-attractive, cubic term in the GPE.\ Similarly, the nonlinear Schr%
\"{o}dinger equation (NLSE) with the self-focusing or defocusing cubic term
(alias the Kerr or anti-Kerr one, respectively) models the transmission of
electromagnetic waves in nonlinear optical media \cite{NLS}.

As well as their linear counterparts, the GPE and NLSE include external
potentials, which often feature the DWP symmetry. However, the symmetry of
the GS in models with the self-focusing nonlinearity (i.e., the state
minimizing the energy at a fixed number of particles in the bosonic gas, or
fixed total power of the optical beam in the photonic medium---in both
cases, these are represented by a fixed norm of the respective wave
function) follows the symmetry of the underlying potential structure only as
long as the nonlinearity remains weak enough. A generic effect, which sets
in with the increase of the strength of the nonlinearity, i.e., effectively,
with the increase of the norm, is \textit{spontaneous symmetry breaking}
(SSB). In its simplest form, the SSB in terms of the BEC implies that the
probability to find the boson in one well of the trapping DWP structure is
larger than in the other. This, incidentally, implies that another basic
principle of quantum mechanics, according to which the GS cannot be
degenerate, is no longer valid in nonlinear models of the quantum origin,
such as the GPE: obviously, the SSB which takes place in the presence of the
DWP gives rise to a degenerate pair of two mutually symmetric ground states,
with the maximum of the wave function observed in either potential well. In
terms of optics, the SSB makes the light power trapped in either core of the
DWP-shaped dual-core waveguide larger than in the mate core. Thus, the SSB
is a fundamental effect common to diverse models of the quantum and
classical origin alike, which combine the wave propagation, nonlinear
self-focusing, and symmetric trapping potentials.

It should be stressed that the same nonlinear system with the DWP potential
always admits a symmetric state coexisting with the asymmetric ones;
however, past the onset of the SSB, the symmetric state no longer represents
the GS, being unstable against small symmetry-breaking fluctuations.
Accordingly, in the course of the spontaneous transition from the unstable
symmetric state to a stable asymmetric one, the choice between the two
mutually degenerate asymmetric states is governed by perturbations, which
``push" the self-attractive system to place, at random, the
maximum of the wave function in the left or right potential well.

In systems with the self-defocusing nonlinearity, the ground state is always
symmetric and stable. In this case, the SSB manifests itself in the form of
the spontaneous breaking of the \textit{antisymmetry} of the first excited
state (the spatially odd one, which has exactly one zero of the wave
function, at the central point, in the one-dimensional geometry). The state
with the spontaneously broken antisymmetry also features a zero, which is
shifted from the central position to the left or right, the sign of the
shift being randomly selected by initial perturbations.

Historically speaking, the SSB concept for nonlinear systems of the NLSE
type was, probably, first proposed by E. B. Davies in 1979 \cite{Davies},
although in a rather abstract mathematical form. In that work, a nonlinear
extension of the Schr\"{o}dinger equation for a pair of quantum particles,
interacting via a three-dimensional isotropic potential, was addressed, and
the SSB was predicted in the form of the spontaneous breaking of the GS
rotational symmetry. Another early prediction of the SSB was reported in the
\textit{self-trapping model}, which is based on a system of linearly coupled
ordinary differential equations with self-attractive cubic terms \cite{Scott}%
. The latter publication had brought the concept of the SSB to the attention
of the broad research community.

\begin{figure}[tbp]
\includegraphics[width=5.0in]{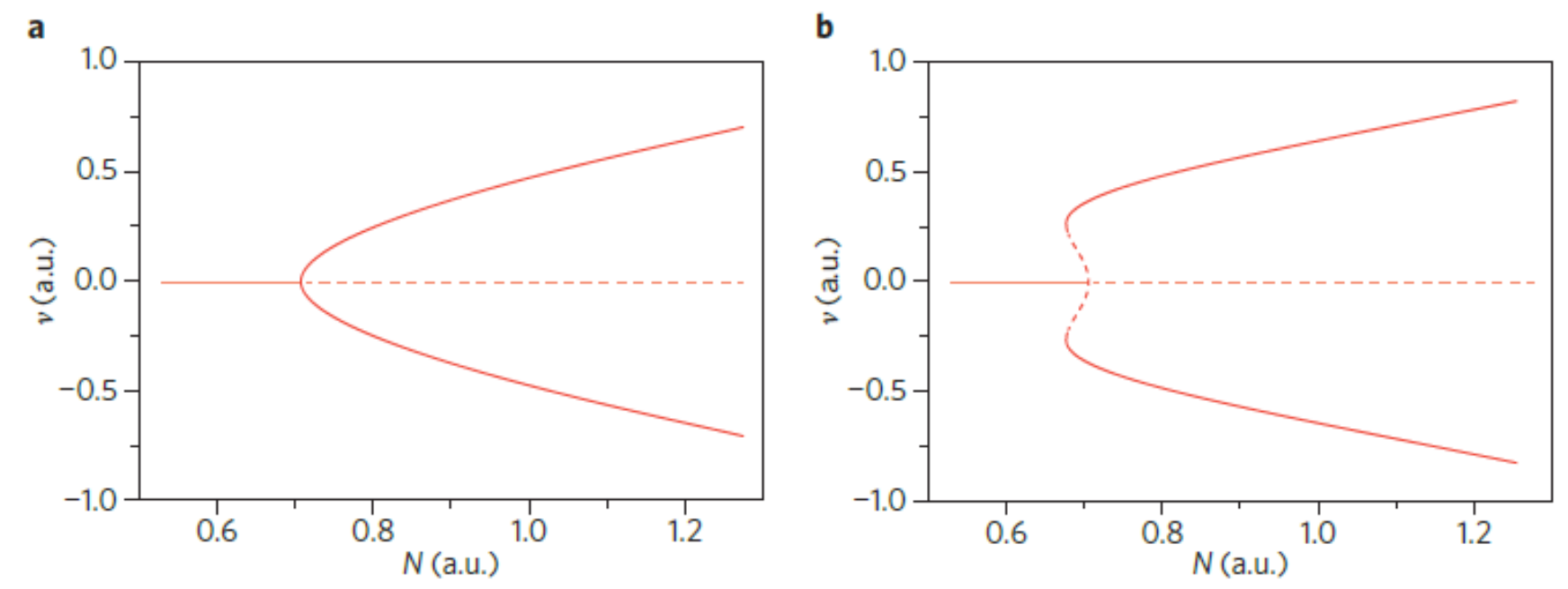}
\caption{{}(Color online) Diagrams for standard supercritical (a) and
subcritical (b) spontaneous-symmetry-breaking bifurcations, as per Ref.
\protect\cite{NatPhot}. Continuous and dashed lines depict, respectively,
stable and unstable solution branches. Total norm $N$ and asymmetry
parameter $\protect\nu $ (see Eq. (\protect\ref{nu})) are shown in arbitrary
units (a.u.).}
\label{fig1}
\end{figure}

An important contribution to theoretical studies of the SSB was made by work
\cite{Snyder}, which addressed this effect in the model for the propagation
of CW (continuous-wave) optical beams in dual-core nonlinear optical fibers
(alias nonlinear directional couplers), with the underlying symmetry between
the linearly coupled cores . In the scaled form, the corresponding system of
propagation equations for CW amplitudes $u_{1}$ and $u_{2}$ in the two cores
is
\begin{eqnarray}
&&i\frac{du_{1}}{dz}+f\left( \left\vert u_{1}\right\vert ^{2}\right)
u_{1}+\kappa u_{2}=0,  \notag \\
&&  \label{u1u2} \\
&&i\frac{du_{2}}{dz}+f\left( \left\vert u_{2}\right\vert ^{2}\right)
u_{2}+\kappa u_{1}=0,  \notag
\end{eqnarray}%
(in this case, ``CW" implies that the amplitudes do not
depend on the temporal variable), where $z$ is the propagation distance, $%
\kappa $ the coefficient accounting for the inter-core linear coupling
through the mutual penetration of evanescent fields from each core into the
mate one, and $f\left( |u_{1,2}|^{2}\right) $ is a function of the intensity
of the light in each core which represents its intrinsic nonlinearity. In
the simplest case of the Kerr (cubic) self-focusing nonlinearity, which
corresponds to%
\begin{equation}
f\left( u^{2}\right) =|u|^{2},  \label{Kerr}
\end{equation}%
this system gives rise to the symmetry-breaking \textit{bifurcation} of the
\textit{supercritical} type \cite{bif}. This bifurcation destabilizes the
symmetric state and, simultaneously, gives rise to a pair of stable
asymmetric ones, which are mirror images of each other, corresponding to
interchange $u_{1}\rightleftarrows u_{2}$, as shown in Fig. \ref{fig1}(a).
In the figure, the asymmetry and the total norm, which characterizes the
strength of the nonlinearity, are defined as%
\begin{equation}
\nu \equiv \left( \left\vert u_{1}\right\vert ^{2}-\left\vert
u_{2}\right\vert ^{2}\right) /\left( \left\vert u_{1}\right\vert
^{2}+\left\vert u_{2}\right\vert ^{2}\right) ,~N\equiv \left( \left\vert
u_{1}\right\vert ^{2}+\left\vert u_{2}\right\vert ^{2}\right) .  \label{nu}
\end{equation}%
On the other hand, the \textit{saturable} nonlinearity, in the form of $%
f\left( \left\vert u\right\vert ^{2}\right) =\left\vert u\right\vert
^{2}/\left( I_{0}+\left\vert u\right\vert ^{2}\right) $, where $I_{0}>0$ is
a constant which determines the intensity-saturation level, gives rise to a
\textit{subcritical} symmetry-breaking bifurcation. In the latter case, the
branches of asymmetric states, which originate at the point of the stability
loss of the symmetric mode, originally evolve \textit{backward} (in terms of
the total power, $\left\vert u_{1}\right\vert ^{2}+\left\vert
u_{2}\right\vert ^{2}$), being unstable, and then turn forward, getting the
stable at the turning point, see Fig. \ref{fig1}(b). This SSB scenario
implies that the pair of stable asymmetric states emerge \textit{%
subcritically}, at a value of the total power smaller than the one at which
the symmetric mode becomes unstable. In terms of statistical physics, the
super- and subcritical bifurcations may be classified as phase transitions
of the second and first kinds, respectively.

The next step in the studies of the SSB phenomenology in models of dual-core
nonlinear optical fibers and similar systems was the consideration of the
fields depending on the temporal variable, $\tau $. In that case, assuming
the anomalous sign of the group-velocity dispersion in each core of the
fiber, Eqs. (\ref{u1u2}) are replaced by a system of NLSEs with the linear
coupling:%
\begin{eqnarray}
&&i\frac{\partial u_{1}}{\partial z}+\frac{1}{2}\frac{\partial ^{2}u_{1}}{%
\partial \tau ^{2}}+f\left( \left\vert u_{1}\right\vert ^{2}\right)
u_{1}+\kappa u_{2}=0,  \notag \\
&&  \label{u1u2tau} \\
&&i\frac{\partial u_{2}}{\partial z}+\frac{1}{2}\frac{\partial ^{2}u_{2}}{%
\partial \tau ^{2}}+f\left( \left\vert u_{2}\right\vert ^{2}\right)
u_{2}+\kappa u_{1}=0.  \notag
\end{eqnarray}%
The same system, with variable $\tau $ replaced by transverse coordinate $x$%
, models the spatial-domain evolution of electromagnetic fields in dual-core
planar waveguides, in which case the second derivatives represent the
paraxial diffraction, instead of the group-velocity dispersion.

The uncoupled NLSEs with the Kerr self-focusing nonlinearity (\ref{Kerr})
give rise to commonly known solitons \cite{NLS}. The corresponding SSB
bifurcation may destabilize obvious symmetric soliton solutions of system (%
\ref{u1u2tau}),%
\begin{equation}
u_{1}=u_{2}=\eta ~\mathrm{sech}\left( \eta \tau \right) \exp \left( \left(
\frac{1}{2}\eta ^{2}+\kappa \right) z\right) ,  \label{soliton}
\end{equation}%
where $\eta $ is an arbitrary real amplitude of the soliton. The bifurcation
replaces the symmetric soliton mode (\ref{soliton}) by asymmetric
two-component modes. The critical value of the soliton's peak power, $\eta
^{2}$, at which the SSB instability of the symmetric solitons sets in under
the action of the Kerr nonlinearity was found in an exact form, $\eta _{%
\mathrm{crit}}^{2}=4/3$, in Ref. \cite{Wabnitz}. The transition to
asymmetric solitons, following the instability onset, was first predicted,
by means of the variational approximation, in Refs. \cite{Pare} and \cite%
{Maim}. Then, it was found that, on the contrary to the supercritical
bifurcation of the CW states in system (\ref{u1u2}) with the Kerr
self-focusing nonlinearity, the SSB bifurcation of the symmetric soliton in
system (\ref{u1u2tau}) is subcritical \cite{Akhmed,Pak}.

An independent line of the analysis of the SSB had originated from the
studies of GPE-based models for atomic BECs trapped in DWP structures. The
scaled form of the corresponding GPE for the mean-field wave function, $\psi
\left( x,t\right) $, is%
\begin{equation}
i\frac{\partial \psi }{\partial t}=-\frac{1}{2}\frac{\partial ^{2}\psi _{1}}{%
\partial x^{2}}-g\left\vert \psi \right\vert ^{2}\psi +U(x)\psi ,
\label{GPE}
\end{equation}%
where $g<0$ and $g>0$ correspond to the repulsive and attractive
collision-induced nonlinearity, respectively. The DWP can be taken, for
instance, as%
\begin{equation}
U(x)=U_{0}\left( x^{2}-a^{2}\right) ^{2},  \label{DWP}
\end{equation}%
with positive constants $U_{0}$ and $a^{2}$.

The GPE (\ref{GPE})\ can be reduced to the two-mode system, similar to the
system of Eqs. (\ref{u1u2}) (with $z$ replaced by $t$), by means of the
tight-binding approximation \cite{tight}, which adopts $\psi (x,t)$ in the
form of a superposition of two stationary wave functions, $\phi $,
corresponding to the states trapped separately in the two potential wells,
centered at $x=\pm a$:%
\begin{equation}
\psi \left( x,t\right) =u_{1}(t)\phi \left( x-a\right) +u_{2}\phi \left(
x+a\right) .  \label{12}
\end{equation}%
In particular, this approximation implies that the nonlinearity acts on each
amplitude $u_{1}$ and $u_{2}$ also separately, while the coupling between
them is linear.

\begin{figure}[tbp]
\includegraphics[width=3.2in]{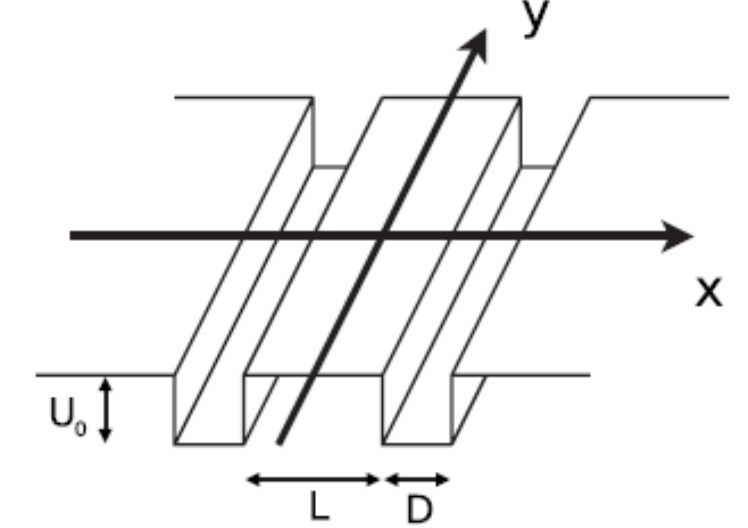}
\caption{{}An example of the quasi-one-dimensional double-trough potential,
built of two parallel potential toughs with the rectangular profile, as per
Ref. \protect\cite{Warsaw}.}
\label{fig2}
\end{figure}

The analysis of the SSB in the models based on the GPE (\ref{GPE}) was
initiated in Refs. \cite{Milburn} and \cite{Smerzi}. Most often, the BEC\
nonlinearity (on the contrary to the self-focusing Kerr terms in optics) is
self-repulsive, which, as mentioned above, gives rise to the spontaneous
breaking of the antisymmetry of the odd states, with $\psi (-x)=-\psi (x)$,
while the GS remains symmetric. Further, the GPE may be extended by adding
an extra spatial coordinate, on which the DWP does not depend, i.e., one
arrives at a two-dimensional GPE with a quasi-one-dimensional\textit{\
double-trough potential}, which is displayed in Fig. \ref{fig2}. In the
latter case, the self-attractive nonlinearity (which, although being less
typical in BEC, is possible too) gives rise to bright matter-wave solitons,
which self-trap in the free direction \cite{soliton}. Accordingly, bright
symmetric solitons are possible in the double-trough potential, and they are
replaced, via a subcritical SSB bifurcation, by stable asymmetric ones at a
critical value of the total norm of the mean-field wave function (which
determines the effective strength of the self-attractive nonlinearity) \cite%
{Warsaw}.

The above discussion was focused on static symmetric and asymmetric modes in
the nonlinear systems featuring the DWP\ structure. The analysis of
dynamical regimes, usually in the form of oscillations of the norm of the
wave function between two wells of the DWP, i.e., roughly speaking, between
the two equivalent asymmetric states existing above the SSB point, has been
developed too. Following the straightforward analogy with Josephson
oscillations of the wave function in tunnel-coupled superconductors \cite%
{superconductor,Ustinov}, the possibility of the matter-wave oscillations in
\textit{bosonic Josephson junctions} was predicted \cite{junction}.

Similar to the situation in many other areas of nonlinear science, the
variety of theoretically predicted results concerning the SSB phenomenology
by far exceeds the number of experimental findings. Nevertheless, some
manifestations of the SSB have been reported in experiments. In particular,
the self-trapping of a macroscopically asymmetric state of the atomic
condensate of $^{87}$Rb atoms with repulsive interactions between them,
loaded into the DWP (which may be considered, as mentioned above, as a
spontaneous breaking of the antisymmetry of the lowest excited state, above
the symmetric GS) and Josephson oscillations in the same system, were
reported in Ref. \cite{Markus}. On the other hand, the SSB of laser beams
coupled into an effective transverse DWP created in a self-focusing
photorefractive medium (where the nonlinearity is saturable, rather than
strictly cubic) has been demonstrated in Ref. \cite{photo}. Still another
experimental observation of the SSB effect in nonlinear optics was the
spontaneously established asymmetric regime of operation of a symmetric pair
of coupled lasers \cite{lasers}. More recently, symmetry breaking was
experimentally demonstrated in a symmetric pair of nanolaser cavities
embedded into a photonic crystal \cite{France} (although the latter system
is a dissipative one, hence its model is essentially different from those
outlined above, cf. Ref. \cite{Sigler}, where the SSB effect for dissipative
solitons was formulated in terms of linearly-coupled complex Ginzburg-Landau
equations with the cubic-quintic nonlinearity). An observation of a related
effect of the spontaneous breaking of the chiral symmetry in metamaterials
was reported in Ref. \cite{Kivshar}.

Many results for the SSB phenomenology and related Josephson oscillations,
chiefly theoretical ones, but also experimental, obtained in various areas
of physics (nonlinear optics and plasmonics, BECs, superconductivity, and
others) are represented by a collection of articles published in topical
volume \cite{book}.

\section{A Simple Model for the Spontaneous Symmetry Breaking (SSB) in a
Double-Well Potential (DWP)}

\subsection{Formulation of the model}

The objective of this section is to introduce what may be the simplest model
which admits the SSB in a system combining the self-attractive nonlinearity
and a DWP structure. In a sketchy form, the model was mentioned in Ref. \cite%
{NatPhot}, but it was not elaborated there. The account given here is not
complete either, as only approximate analytical results are included. A full
presentation, including relevant numerical results, will be given elsewhere.

\begin{figure}[tbp]
\includegraphics[width=3.2in]{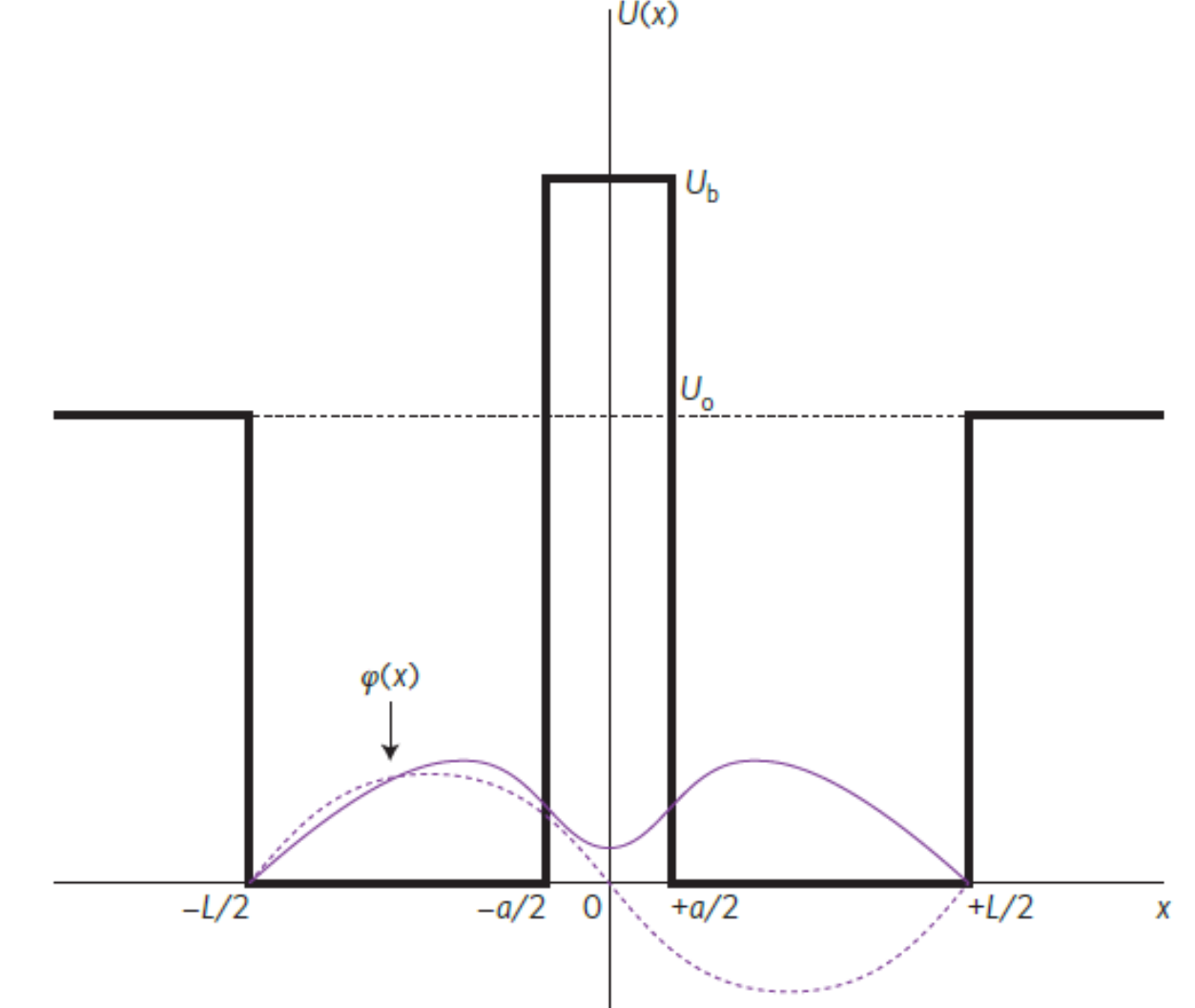}
\caption{{}(Color online) A sketch of the double-well-potential (DWP)
structure under the consideration (as per Ref. \protect\cite{NatPhot}): an
infinitely deep potential box ($U_{0}\rightarrow \infty $), of width $%
L\equiv 1$, is split in the middle by a narrow tall barrier, $\protect%
\varepsilon \protect\delta (x)$, see Eq. (\protect\ref{NLSE}). Even and odd
wave functions of the ground and first excited states, in the absence of the
spontaneous symmetry breaking, are shown by the continuoys and dashed
curves, respectively.}
\label{fig3}
\end{figure}

The model is schematically shown in Fig. \ref{fig3}. It is built as an
infinitely deep potential box, with the DWP structure created by means of
the delta-functional barrier created in the center, cf. the cross section of
the double-trough potential displayed in Fig. \ref{fig2}. The respective
scaled form of the GPE is given by Eq. (\ref{GPE}) with $g>0$ and $%
U(x)=\varepsilon \delta (x)$, where the delta-functional potential
corresponds to $U_{b}\rightarrow \infty ,$ $a\rightarrow 0$, while the
strength of the barrier, $\varepsilon \equiv U_{b}a,$ is kept fixed. The
edges of the potential box at points $x=\pm 1/2$ are represented by the
boundary conditions (b.c.)%
\begin{equation}
\psi \left( x=\pm \frac{1}{2}\right) =0.  \label{bc}
\end{equation}%
Stationary states with chemical potential $\mu $ are looked for as $\psi
(x,t)=e^{-i\mu t}\phi (x),$ with real function $\phi (x)$ obeying the
following stationary equation with the respective b.c.:
\begin{equation}
\mu \phi =-\frac{1}{2}\frac{d^{2}\phi }{dx^{2}}-g\phi ^{3}+\varepsilon
\delta (x)\phi ,~\phi \left( x=\pm \frac{1}{2}\right) =0.  \label{NLSE}
\end{equation}%
The delta-functional barrier at $x=0$ implies that $\phi (x)$ is continuous
at this point, while its derivative features a jump:%
\begin{equation}
\frac{d\phi }{dx}|_{x=+0}-\frac{d\phi }{dx}|_{x=-0}=2\varepsilon \phi (x=0).
\label{jump}
\end{equation}%

It is possible to fix $g\equiv 1$ in Eqs. (\ref{GPE}) and (\ref{NLSE}) by
means of scaling, but it is more convenient, for the sake of the subsequent
analysis, to keep $g>0$ as a free parameter. The strength of the
nonlinearity is determined by product $gN$, where the total norm of the wave
function is defined as the sum of the norms trapped in the left and right
potential wells (cf. Eq. (\ref{nu})):%
\begin{equation}
N=\left( \int_{-1/2}^{0}+\int_{0}^{+1/2}\right) \phi ^{2}(x)dx\equiv
N_{-}+N_{+},  \label{N}
\end{equation}

Before proceeding to the analysis of the SSB in the nonlinear model, it is
relevant to briefly discuss its linear counterpart, with $g=0$ in Eq. (\ref%
{NLSE}). Spatially symmetric (even) solutions of this equation are looked for
as%
\begin{equation}
\phi _{\mathrm{even}}^{(\mathrm{lin})}(x)=A\sin \left( \sqrt{2\mu }\left(
\frac{1}{2}-|x|\right) \right) ,  \label{linear}
\end{equation}%
where $A$ is an arbitrary amplitude, and eigenvalue $\mu $ is determined by
the equation following from the jump condition (\ref{jump}):%
\begin{equation}
\tan \left( \sqrt{\mu /2}\right) =-\sqrt{2\mu }/\varepsilon .  \label{eigen}
\end{equation}%
It is easy to see that, with the increase of $\varepsilon $ from $0$ to $%
\infty $, $\mu _{0}$ the lowest eigenvalue $\mu _{0}$, corresponding to the
GS of the linear model, monotonously grows from $\mu _{0}\left( \varepsilon
=0\right) =\pi ^{2}/2$ to
\begin{equation}
\mu _{0}\left( \varepsilon =\infty \right) =2\pi ^{2}.  \label{muGS}
\end{equation}%
Similarly, the eigenvalue of the first excited symmetric state, $\mu
_{2}\left( \varepsilon \right) $, monotonously grows from to $\mu _{2}\left(
\varepsilon =0\right) =9\pi ^{2}/2$ to $\mu _{2}\left( \varepsilon =\infty
\right) =8\pi ^{2}$. Located between eigenvalues $\mu _{0}$ and $\mu _{2}$,
is $\mu _{1}=2\pi ^{2},$ which corresponds to the lowest excited state,
i.e., the first antisymmetric (spatially odd) eigenfunction, $\phi _{\mathrm{%
odd}}^{(\mathrm{lin})}(x)=A\sin \left( \sqrt{2\mu _{1}}x\right) $.
Naturally, $\mu _{1}$ coincides with the limit value (\ref{muGS}) of $\mu
_{0}$, and it does not depend on $\varepsilon $, as the odd eigenfunction
vanishes at $x=0$, where the $\delta $-function is placed.

\subsection{An analytical solution for the SSB point in the strongly-split
DWP (large $\protect\varepsilon $)}

The main objective of the analysis is to predict the critical norm which
gives rise to the SSB, through the competition between the self-focusing,
which favors the spontaneous accumulation of the wave function in one well,
and the linear coupling between the wells, which tends to distribute the
wave function evenly between them. An approximate analytical solution to
this problem can be obtained in the case of weakly coupled potential wells,
which corresponds to large $\varepsilon $ (a very tall central barrier). In
this case, weak nonlinearity, i.e., a small amplitude of the wave function,
is sufficient to induce the SSB. In turn, the small amplitude implies that
solutions to Eq. (\ref{NLSE}) vanishing at $x=\pm 1/2$ are close to
eigenmodes (\ref{linear}) of the linearized version of the same equation,
i.e., the approximate solutions may be sought for as%
\begin{equation}
\phi (x)=A_{\pm }\sin \left( k_{\pm }\left( \frac{1}{2}-|x|\right) \right) ,
\label{lin}
\end{equation}%
where signs $\pm $ pertain to the regions of $x<0$ and $x>0$, respectively, $%
k_{\pm }$ being appropriate wavenumbers. The substitution of \textit{ansatz}
(\ref{lin}) into the condition of the continuity of the wave function at $x=0
$, and the jump condition (\ref{jump}) for the first derivatives, yields the
following relations between amplitudes $A_{\pm }$ and the wavenumbers:%
\begin{gather}
A_{+}\sin \left( \frac{1}{2}k_{+}\right) =A_{-}\sin \left( \frac{1}{2}%
k_{-}\right) ,  \label{bc1} \\
A_{-}k_{-}\cos \left( \frac{1}{2}k_{-}\right) -A_{+}k_{+}\cos \left( \frac{1%
}{2}k_{+}\right) =4\varepsilon A_{\pm }\sin \left( \frac{1}{2}k_{\pm
}\right) .  \label{bc2}
\end{gather}%
Further, in the same small-amplitude limit, the cubic term in Eq. (\ref{NLSE}%
) may be approximated by the neglecting the third harmonic contained in it:%
\begin{equation}
\left[ A_{\pm }\sin \left( k_{\pm }\left( \frac{1}{2}-|x|\right) \right) %
\right] ^{3}\approx \frac{3}{4}A_{\pm }^{3}\sin \left( k_{\pm }\left( \frac{1%
}{2}-|x|\right) \right) ,  \label{3/4}
\end{equation}%
which, in turn, implies an effective shift of the chemical potential in Eq. (%
\ref{NLSE}) and determines the corresponding wavenumbers in Eq. (\ref{lin}):%
\begin{equation}
k_{\pm }=\sqrt{2\left( \mu +\frac{3}{4}gA_{\pm }^{2}\right) }.  \label{k}
\end{equation}

In the limit of $\varepsilon \rightarrow \infty $, wave functions (\ref{lin}%
) must vanish at $x=0$, hence the respective GS corresponds to $k_{\pm
}=2\pi $, i.e., to the above-mentioned value (\ref{muGS}) of the chemical
potential. In the same limit, the norm (\ref{N}) of the GS is
\begin{equation}
N=\left( A_{-}^{2}+A_{+}^{2}\right) /4.
\end{equation}%
At large but finite $\varepsilon $, the GS chemical potential is sought for
as%
\begin{equation}
\mu =2\pi ^{2}-\delta \mu ,~\mathrm{with~~}\delta \mu \ll 2\pi ^{2}.
\label{delta-mu}
\end{equation}%
Next, the substitution of this expression into Eq. (\ref{k}), expanding it
for small $\delta \mu $ and $A_{\pm }^{2}$, and inserting the result into
Eqs. (\ref{bc1}) and (\ref{bc2}) leads to equations which take a relatively
simple form at the point of the onset of the SSB bifurcation, i.e., in the
limit of $A_{+}-A_{-}\rightarrow 0$ (the vanishingly small factor\ $\left(
A_{+}-A_{-}\right) $ then factorizes out and cancels in in the expanded
version of Eq. (\ref{bc1})):%
\begin{equation}
N_{\mathrm{cr}}=\frac{8\pi ^{2}}{3g}\varepsilon ^{-1},~\left( A_{\pm
}^{2}\right) _{\mathrm{cr}}=2N_{\mathrm{cr}},~\delta \mu =12\pi
^{2}\varepsilon ^{-1}.  \label{cr}
\end{equation}%
This result was mentioned, without the derivation, in Ref. \cite{NatPhot}.
Thus, as expected, the critical value of the norm at the SSB point decays ($%
\sim \varepsilon ^{-1}$) with the increase of $\varepsilon $. The
substitution of $\delta \mu $ from Eq. (\ref{cr}) into Eq. (\ref{delta-mu})
suggests that this asymptotic solution is actually valid for $\varepsilon
\gg 6$.

\subsection{An analytical solution for the SSB in the weakly-split DWP
(small $\protect\varepsilon $): the soliton approximation}

The case of small $\varepsilon $, opposite to that considered above, implies
that the central barrier splitting the confined box into the two potential
wells is weak, hence the effective coupling between the wells is strong.
According to the general principles of the SSB theory \cite{book}, strong
nonlinearity, i.e., large norm $N$, is necessary to complete with the strong
coupling. Large $N$, in turn, implies that the wave field self-traps into a
narrow NLSE\ soliton \cite{NLS},%
\begin{equation}
\phi _{\mathrm{sol}}\left( x-\xi \right) =\frac{1}{2}\sqrt{g}N\mathrm{sech}%
\left( \frac{g}{2}N\left( x-\xi \right) \right) ,  \label{sol}
\end{equation}%
where $\xi $ is the coordinate of the soliton's center, the respective
chemical potential is%
\begin{equation}
\mu _{\mathrm{sol}}=-\left( gN\right) ^{2}/8.  \label{mu-sol}
\end{equation}%
and it is assumed that $N$ is large enough to make the soliton's width much
smaller than the size of the box ($L\equiv 1$ in Fig. \ref{fig3}), i.e.,
\begin{equation}
gN\gg 1.  \label{>>}
\end{equation}

The soliton is repelled from the edges of the potential box. To comply with
the b.c. in Eq. (\ref{NLSE}), this may be interpreted as the repulsive
interaction with two \textit{ghost solitons} generated as mirror images
(with opposite signs) of the real one (\ref{sol}) with respect to the edges
of the box:%
\begin{equation}
\phi _{\mathrm{ghost}}(x)=-\sqrt{g}\left( N/2\right) \left[ \mathrm{sech}%
\left( \frac{g}{2}N\left( x-\frac{1}{2}+\xi \right) \right) +\mathrm{sech}%
\left( \frac{g}{2}N\left( x+1+\xi \right) \right) \right] .  \label{ghost}
\end{equation}%
The potential of the interaction between two far separated NLSE solitons is
well known \cite{sol-sol1}-\cite{sol-sol4}. In the present case, the sum of
the two interaction potentials, corresponding to the pair of the ghosts,
amounts to the following effective potential accounting for the repulsion of
the real soliton from edges of the box in which it is confined:%
\begin{equation}
U_{\mathrm{box}}(\xi )=g^{2}N^{3}\exp \left( -\frac{g}{2}N\right) \cosh
\left( Ng\xi \right) .  \label{box}
\end{equation}

On the other hand, the soliton is repelled by the central barrier, the
respective potential being \cite{RMP}%
\begin{equation}
U_{\mathrm{barrier}}(\xi )=\varepsilon \phi _{\mathrm{sol}}^{2}\left( \xi
=0\right) =\frac{\varepsilon g}{4}N^{2}\mathrm{sech}^{2}\left( \frac{g}{2}%
N\xi \right) ,  \label{barrier}
\end{equation}%
where the latter expression was obtained neglecting the deformation of the
soliton's shape. A straightforward analysis of the total effective
potential, $U(\xi )=U_{\mathrm{box}}(\xi )+U_{\mathrm{barrier}}(\xi )$,
demonstrates that the position of the soliton placed at $\xi =0$, which
represents the symmetric state in the present case, is stable, i.e., it
corresponds to a minimum of the net potential, at%
\begin{equation}
8gN\exp \left( -\frac{g}{2}N\right) >\varepsilon ,  \label{>}
\end{equation}%
or, in other words, at%
\begin{equation}
N>N_{\mathrm{cr}}\approx \left( 2/g\right) \ln \left( 16/\varepsilon \right)
\label{Ncr}
\end{equation}%
(the underlying assumption that $\varepsilon $ is small was used to derive
Eq. (\ref{Ncr}) from Eq. (\ref{>})). With the increase of $N,$ the SSB
bifurcation takes place at $N=N_{\mathrm{cr}}$, when the potential minimum
at $\xi =0$ turns into a local maximum. At $0<\left( N-N_{\mathrm{cr}%
}\right) /N_{\mathrm{cr}}\ll 1$, the center of the soliton spontaneously
shifts to either of two asymmetric positions, which correspond to a pair of
emerging potential minima at $\xi \neq 0$:%
\begin{equation}
\xi =\pm \sqrt{\left( N-N_{\mathrm{cr}}\right) /\left( gN_{\mathrm{cr}%
}^{2}\right) }.  \label{xi}
\end{equation}%
The latter result explicitly describes the SSB bifurcation of the
supercritical type, which occurs in the present setting.

\subsection{The variational approximation for the SSB in the generic DWP}

A possibility to develop a more comprehensive, albeit coarser, analytical
approximation for solutions of Eq. (\ref{NLSE}) in the generic case (when the
strength of the splitting barrier, $\varepsilon $, is neither specifically
large nor small) is suggested by the variational approximation (VA) \cite%
{Progress}. To this end, note that Eq. (\ref{NLSE}) can be derived from the
Lagrangian,%
\begin{equation}
L=\int_{-1/2}^{+1/2}\mathcal{L}(x)dx,~\mathcal{L}=\frac{1}{2}\left( \frac{%
d\phi }{dx}\right) ^{2}-\mu \phi ^{2}-\frac{g}{2}\phi ^{4}+\varepsilon
\delta (x)\phi ^{2}.  \label{L}
\end{equation}

Aiming to apply the VA for detecting the SSB onset point, one can adopt the
following \textit{ansatz} for the GS wave function:

\begin{equation}
\phi (x)=a\cos (\pi x)+b\sin (2\pi x)+c\cos (3\pi x),  \label{ans}
\end{equation}%
with each term satisfying b.c. in Eq. (\ref{NLSE}). Real constants $a,$ $c,$
and $b$ must be predicted by the VA. The SSB is accounted for by terms $\sim
b$ in ansatz (\ref{ans}), hence the onset of the SSB is heralded by the
emergence of a solution with infinitesimal $b$, branching off from from the
symmetric solution with $b=0$, similar to how the onset of the SSB\
bifurcation in terms of ansatz (\ref{lin}) is signaled by the emergence of
infinitesimal $\left( A_{+}-A_{\_}\right) $ in the above analysis.

The total norm (\ref{N}) of ansatz (\ref{ans}) is $N=\left( 1/2\right)
\left( a^{2}+c^{2}+b^{2}\right) $, while its asymmetry at $b\neq 0$ may be
quantified by%
\begin{equation}
\Theta \equiv \frac{N_{+}-N_{-}}{N}=\frac{16}{15\pi }\frac{b\left(
5a-3c\right) }{a^{2}+c^{2}+b^{2}}.  \label{Theta}
\end{equation}%
A straightforward consideration demonstrates that, for all values of $a,b,$
and $c$, expression (\ref{Theta}) is subject to constraint $\left\vert
\Theta \right\vert <1$, as it must be. When $b=0$, the theorem that the
spatially symmetric GS cannot have nodes, i.e., $\phi (x)\neq 0$ at $|x|<1/2$%
, if applied to ansatz (\ref{ans}), easily amounts to the following
constraint:%
\begin{equation}
-1<c/a<1/3.  \label{or}
\end{equation}

Further, the substitution of ansatz (\ref{ans}) into Lagrangian (\ref{L})
yields

\begin{gather}
L=\allowbreak \left( \frac{1}{4}\pi ^{2}-\frac{1}{2}\mu +\varepsilon \right)
a^{2}+\left( \pi ^{2}-\frac{1}{2}\mu \right) b^{2}+\left( \frac{9}{4}\pi
^{2}-\frac{1}{2}\mu +\varepsilon \right) c^{2}\allowbreak +2\varepsilon ac
\notag \\
-\frac{g}{4}\left( \frac{3}{4}a^{4}+a^{3}c+3a^{2}b^{2}+3a^{2}c^{2}-3ab^{2}c+%
\frac{3}{4}\allowbreak b^{4}+3b^{2}c^{2}+\frac{3}{4}c^{4}\right) ,
\label{LL}
\end{gather}%
from which three variational equations follow:
\begin{equation}
\partial L/\partial (b^{2})=0,  \label{d/db}
\end{equation}%
\begin{equation}
\partial L/\partial a=\partial L/\partial c=0.  \label{dd/dadc}
\end{equation}%
In the general form, these equations are rather cumbersome. However, being
interested in the threshold at which the SSB sets in, one may set $b=0$ in
these equations (after performing the differentiation with respect to $b^{2}$
in Eq. (\ref{d/db})), which lead to the following system of three equations
for three unknowns $a$, $b$, and $\mu $:
\begin{equation}
2\pi ^{2}-\mu =\frac{3g}{2}\left( a^{2}-ac+c^{2}\right) ,  \label{bsimple}
\end{equation}%
\begin{gather}
\left( \frac{1}{2}\pi ^{2}-\mu +2\varepsilon \right) a+2\varepsilon c-\frac{g%
}{4}\left( 3a^{3}+3a^{2}c+6ac^{2}\right) =0,  \label{asimple} \\
\left( \frac{9}{2}\pi ^{2}-\mu +2\varepsilon \right) c\allowbreak
+2\varepsilon a-\frac{g}{4}\left( a^{3}+6a^{2}c+3c^{3}\right) =0.
\label{csimple}
\end{gather}

In particular, Eqs. (\ref{asimple}) and (\ref{csimple}) with $g=0$, while
Eq. (\ref{bsimple}) is dropped, offer an additional application: they
predict the GS chemical potential of the linear system ($g=0$), as the value
at which the determinant of the linearized version of Eqs. (\ref{asimple})
and (\ref{csimple}) for $\ a$ and $c$ vanishes:
\begin{equation}
\mu _{0}=\frac{5}{2}\pi ^{2}+2\varepsilon -2\sqrt{\pi ^{4}+\varepsilon ^{2}}
\label{GS}
\end{equation}%
(recall that $\mu _{0}(\varepsilon )$ cannot be found above in an exact
form). The latter approximation is meaningful once it yields the GS chemical
potential smaller than the above-mentioned exact value $2\pi^2$ corresponding
to the lowest excited state. This condition holds at%
\begin{equation}
\varepsilon <\left( 15/8\right) \pi ^{2}\approx 18.5.  \label{<}
\end{equation}%
Further, it is easy to check that Eqs. (\ref{bsimple})-(\ref{csimple}) yield
no physical solutions at $\varepsilon =0$, in agreement with the obvious
fact that the SSB does not occur when the central barrier is absent, i.e.,
the potential is not split into two wells.

Finally, a particular exact solution of Eqs. (\ref{bsimple})-(\ref{csimple})
(which includes a particular value of $\varepsilon $) can be found by
setting $c=0$, i.e., assuming that the third harmonic vanishes in ansatz (%
\ref{ans}):%
\begin{equation}
a^{2}=3\pi ^{2}/\left( 2g\right) ,~\mu =-\pi ^{2}/4,~\varepsilon =3\pi
^{2}/16\approx 1.85.  \label{b=c=0}
\end{equation}%
A noteworthy feature of this particular solution is that it has $\mu <0$.
Indeed, Eq. (\ref{NLSE}) suggests that a sufficiently strong nonlinear term
should make the chemical potential negative, as corroborated by Eq. (\ref%
{mu-sol}).

A consistent analysis of the VA for the present model, and its comparison
with numerical results will be reported elsewhere.

\section{Conclusion}

The objective of this paper was two-fold. First, a short overview was given
of the general topic of the SSB (spontaneous symmetry breaking) in nonlinear
one-dimensional models featuring the competition of the self-focusing cubic
nonlinearity and DWP (double-well-potential) structure. Physically relevant
examples of such systems are offered by nonlinear optical waveguides with
the transverse DWP structure, and by BEC trapped in two symmetric potential
wells coupled by tunneling of atoms. The SSB occurs at a critical value of
the nonlinearity strength, i.e., of the field's norm (which is tantamount to
the total power, in the case of the trapped optical beam).

The second part of the paper displayed a particular model, which is the
simplest one capable to grasp the SSB phenomenology: an infinitely deep
potential box, split into two wells by a delta-functional barrier set at the
center. Approximate analytical results predicting the SSB point have been
presented for two limit cases, \textit{viz}., the strong or weak split of
the potential box by the central barrier. In both cases, critical values of
the norm at the SSB point have been found, being, respectively, small and
large. For the generic (intermediate) case, a coarser approach based on the
VA (variational approximation) has been developed. The detailed analysis of
the VA and comparison of the predictions with numerical results will be
reported elsewhere.

\end{document}